\documentclass[a4paper]{article}

\usepackage{INTERSPEECH2019}
\usepackage{amsmath,graphicx}
\usepackage{xargs}                      

\usepackage[pdftex,dvipsnames]{xcolor}  
\usepackage[colorinlistoftodos,prependcaption,textsize=tiny]{todonotes}
\usepackage{hyperref}
\usepackage{tabularx}

\usepackage{bm}
\usepackage{amssymb,amsmath,graphicx,bbm,color,booktabs}
\usepackage{amsopn}

\usepackage{xcolor}
\usepackage{xargs}                      
\usepackage[pdftex,dvipsnames]{xcolor}  
\usepackage{todonotes}
\newcommand{\addref}[1]{\todo[color=red!40]{Add reference.}}

\usepackage[colorinlistoftodos,prependcaption,textsize=tiny]{todonotes}

\newcommand{\vw}{\bm{w}}

\newcommand{\vphi}    {\bm{\phi}}

    \newcommand{\Yc}{\mathcal{Y}}

\newcommand{\R}{\mathbb{R}}

\newcommand{\reals}{\mathbb{R}}

\DeclareMathOperator*{\argmin}{argmin}
\DeclareMathOperator*{\argmax}{argmax}

\newcommand{\X}{\mathcal{X}}
\newcommand{\sx}{\bar{\x}}

\newcommand{\sy}{\bar{\y}}
\newcommand{\syh}{\sy'}
\newcommand{\tb}{t_b}

\newcommand{\tv}{t_v}
\newcommand{\tpv}{t_{pv}}

\newcommand{\T}{\mathcal{T}}

\newcommand{\tableref}[1]{Table~\ref{#1}}

\renewcommand{\eqref}[1]{Eq.~(\ref{#1})}

\def \x{{\mathbf x}}
\def \y{{\mathbf y}}

\newcommand{\figref}[1]{Fig.~\ref{#1}}

\title{Dr.VOT : Measuring Positive and Negative Voice Onset Time in the Wild}
\name{Yosi Shrem$^1$, Matthew Goldrick$^{2}$, Joseph Keshet$^{1}$}

\address{$^1$Department of Computer Science, Bar-Ilan University, Ramat-Gan, Israel\\
	$^2$Department of Linguistics, Northwestern University, Evanston, IL, USA}
\email{jkeshet@cs.biu.ac.il} 

\begin{document}
	
	\maketitle

\begin{abstract}
Voice Onset Time (VOT), a key measurement of speech for basic research and applied medical studies, is the time between the onset of a stop burst and the onset of voicing. When the voicing onset precedes burst onset the VOT is negative; if voicing onset follows the burst, it is positive. In this work, we present a deep-learning model for accurate and reliable measurement of VOT in naturalistic speech. The proposed system addresses two critical issues: it can measure positive and negative VOT equally well, and it is trained to be robust to variation across annotations. Our approach is based on the structured prediction framework, where the feature functions are defined to be RNNs. These learn to capture segmental variation in the signal. Results suggest that our method substantially improves over the current state-of-the-art. In contrast to previous work, our Deep and Robust VOT annotator, Dr.VOT, can successfully estimate negative VOTs– while maintaining state-of-the-art performance on positive VOTs. This high level of performance generalizes to new corpora without further retraining. 
\end{abstract}
\noindent\textbf{Index Terms}: structured prediction, multi-task learning, adversarial training, recurrent neural networks, sequence segmentation.

\section{Introduction}

In acoustic studies of speech, one of the most commonly-examined properties of sounds is the voicing of stop consonants (e.g., the contrast between English /b d g/ and /p t k/). In initial positions this is typically indexed by voice onset time (VOT), the time between the release burst of the obstruent and the onset of voicing (negative when voicing begins before the release, and positive when voicing follows release) \cite{lisker1964cross}. The VOT task makes major contributions to both clinical \cite{auzou2000voice} and theoretical \cite{Paterson2011} studies, for example: to characterize how a communication disorder affects speech \cite{auzou2000voice}, or how languages differ in phonetic cues to stop contrasts\cite{lisker1964cross,cho1999variation}. 

Being a key feature in distinguishing voiced and voiceless consonants across languages, VOT is increasingly used as a feature for ASR tasks, such as stop consonant classification \cite{niyogi2003voicing,stouten2009automatic-short,hansen2010automatic}.
A core challenge to the use of VOT in acoustic studies is the need for highly accurate measurements. For example, speakers use small variations in VOT to signal contrasts with a similar word (e.g., signaling that the speaker said pill instead of bill \cite{baese2009mechanisms}). Because such effects are modulated across different contexts \cite{buz2016dynamically}, such accurate measurements must also be scalable. These issues have been partially addressed in our previous work using discriminative large-margin algorithms to measure positive VOT, trained on data from human expert coders \cite{sonderegger2012automatic,goldrick2016automatic}. 
This approach has a significant shortcoming in failing to address the measurement of negative VOTs, severely limiting the application of this tool. While such VOTs are relatively rare in English, they are the prototypical realization of voiced stops in many other languages (e.g., many Romance and Semitic languages). The amplitude of voicing occurring before release of closure (as in negative VOTs) can be quite small and highly variable, reflecting the challenges of maintaining voicing during closure \cite{abramson2017voice}. Variation of such weak signals can lead to substantial variation in the application of annotations. Extending algorithms \cite{adi2015vowel,adi2017sequence} to cover such instances has been challenging, in part, due to variation in the annotation. While there is some variation in the annotation of positive VOT \cite{sonderegger2012automatic}, the annotations of negative VOT is extensively diversified. Hence, applying the same measurement is inadequate. A related challenge is the varying structure of different datasets. As noted above, the distribution of VOTs varies across languages, and, within a single language community, there is substantial inter-speaker variation in VOT \cite{sonderegger2017medium}, possibly reflecting systematic differences as a function age, gender, and other sociolinguistic variables \cite{torre2009age}.

Inspired by recent developments in Multi-Task Learning (MTL) \cite{caruana1997multitask, collobert2011natural} and adversarial training techniques \cite{goodfellow2014generative,ganin2016domain,madry2017towards}, we explored these methods separately and combined in a structured prediction framework.
MTL \cite{caruana1997multitask,collobert2011natural} is a technique to enhance overall performance and robustness by learning multiple domain-related tasks. Relying on the assumption that both tasks share a familiar set of underlying features, leveraging information from an auxiliary task can improve the main task accuracy. MTL can focus the model's attention to relevant features that single-task training may overlook. For example, in speech recognition, an auxiliary task can be phonetic context identification or gender classification \cite{pironkov2016multi}. 
Our training data is composed of several corpora. Each was recorded in a different environment, and annotated by different annotators using slightly different criteria. Naturally, characteristics of the speaker such as age, gender, and so on, also vary and are known to influence  VOT \cite{singh2016relationship}. One technique for addressing this issue is adversarial training, which increases model robustness by making it invariant to a specified criterion. To that end, a set of models are trained together but pursue competing goals. Specifically, a good acoustic representation should be corpus-invariant. It should encode only the relevant features regardless of the recording environment, such that a dataset classifier will not be able to distinguish between examples from different corpora. 
Since the acoustic features vary across corpora, current approaches \cite{adi2015vowel,adi2017sequence} are unable to transfer good performance on one dataset to different environment. 

To that end, we propose {\bf Dr.VOT}, a {\bf Deep Robust method for VOT measurement}, incorporating adversarial training and MTL within a structured prediction framework, so as to enhance general performance and directly address the lack of generalization to new environments and domains. In this paper, the classifying of VOT type as positive or negative as an auxiliary task for MTL, while the adversary task is to find the corpus an utterance originated from.

		\vspace{-8pt}
\section{Problem Settings}
VOT measurement belongs to the family of segmentation problems, as we are provided with a series of time-related samples in an arbitrary length, and required to predict the boundaries for every segment. Let $ \bar{\x} = (\x_1,...,\x_T) $ denote the input speech utterance, represented as a sequence of acoustic features. Every frame $\x_t$ for $ t \in [1,T]$ is a D dimensional vector, i.e. $\x_t \in \R^D$. Since the input can be of any length, the number of frames denoted as $T$ is not fixed.

The corresponding segmentation for every input is denoted by $\bar{\y} = (y_1,\ldots,y_k)$, where $k$ can vary across different inputs. Each element $y_i$ in the label vector $\bar{\y}$, indicates the start time of a new segment, hence $y_i \in \T$ where $\T=\{1,\ldots,T\}$, and the number of segments is denoted by $|\bar{\y}|$.

The VOT is defined as the time difference between the burst onset and the voicing onset. When the voice onset precedes the burst, it is called prevoiced, and the VOT is negative. Otherwise, when the voice comes after the burst, the VOT is positive.
In order to measure VOT, each utterance is paired with three indices corresponding to the prevoicing onset, $\tpv \in \T \cup \{-1\}$ (in case of positive VOT, no voice precedes the burst), burst onset, $\tb \in \T$, and the vowel onset $\tv \in \T$, s.t. $\tpv < \tb < \tv$. In the positive case, the VOT is $t_v - t_b$, and the segmentation is $\bar{\y}=(\tb, \tv)$. Similarly, if negative, the VOT is $t_b - t_{pv}$ and the segmentation is $\bar{\y}=(\tpv, \tb)$.




\section{MODEL ARCHITECTURE}
In this section, we explain our model and the joint optimization process thoroughly.
First, we provide basic knowledge about structured prediction. We then describe a basic structured model using an RNN as a feature function, and then combine this with VOT classification and an adversarial branch.
\subsection{Structured Prediction}\label{ssec:structured}
For every pair $(\sx,\sy)$ where $\sx$ is the input sequence, $\sy$ is the desired output. Our goal is to find $\syh_{\vw}$ which is a good approximation to the true output $\sy$:
\begin{equation}
\label{eq:yw}
\syh_{\vw}(\sx) = \argmax_{\sy \in \Yc} ~ \vw^\top \Phi(\sx, \sy),
\vspace{-3pt}
\end{equation}
where $\vw \in \reals^D$ is the set of model parameters, which are estimated on a training set. The feature function $\Phi: \X \times \Yc \rightarrow \reals^D$, which is described in the following section, maps a real-world object to a real values vector. In order to quantify a prediction's quality, we define a {\it task loss} denoted by $\ell$, as the absolute difference between our prediction and the true segmentation:
\begin{equation}
\label{eq:task_loss}
\ell(\sy,\syh) =  \left[|y_1 -  y'_1| - \tau]_+ +[|y_2 - y'_2| - \tau\right]_+,
\vspace{-2pt}
\end{equation} 
where $[\pi]_+=\max\{0,\pi\}$, and $\tau$ is a user defined tolerance parameter. 
In order to find the model parameters $\vw$, ideally we would like to minimize the expected task loss. Let $\rho$ be the unknown data distribution, then:
\begin{equation}\label{eq:reg-loss}
\vw^* = \argmin_{\vw} ~ \mathbb{E}_{(\sx,\sy) \sim \rho} [\ell(\sy,\syh_{\vw}(\sx))].
\vspace{-3pt}
\end{equation}
Since the distribution $\rho$ is unknown, we minimize the empirical loss over a training set 
of examples that are drawn i.i.d. from $\rho$. Moreover, the non-convexity of the loss function makes direct optimization a hard task, hence, we minimize a closely related and slightly different function, called {\it surrogate loss}, denoted $\bar{\ell}_\mathrm{struct}(\vw,\sx,\sy)$ that upper bounds the loss:
\begin{equation}
\label{eq:structured-loss}
\begin{aligned}
\bar{\ell}_\mathrm{struct}(\vw, \sx, \sy) = \max_{\syh \in \Yc} ~ [\ell(\sy,\syh) - \vw^\top\Phi(\sx,\sy) + \vw^\top\Phi(\sx,\syh)] .
\end{aligned}
\end{equation}
\subsection{RNNs as Feature Functions}\label{ssec:rnn}
A recurrent neural network (RNN) is a neural architecture that excels in capturing relations between elements within a temporal sequence. Using an inner state which acts as a memory, an RNN leverages insights from previous frames while predicting the next one. Such architectures have excelled in multiple temporal domains such as NLP \cite{mikolov2010recurrent} and speech \cite{graves2013speech}.

\begin{figure}[t]
	\centering   
	\includegraphics[height=4.2cm]{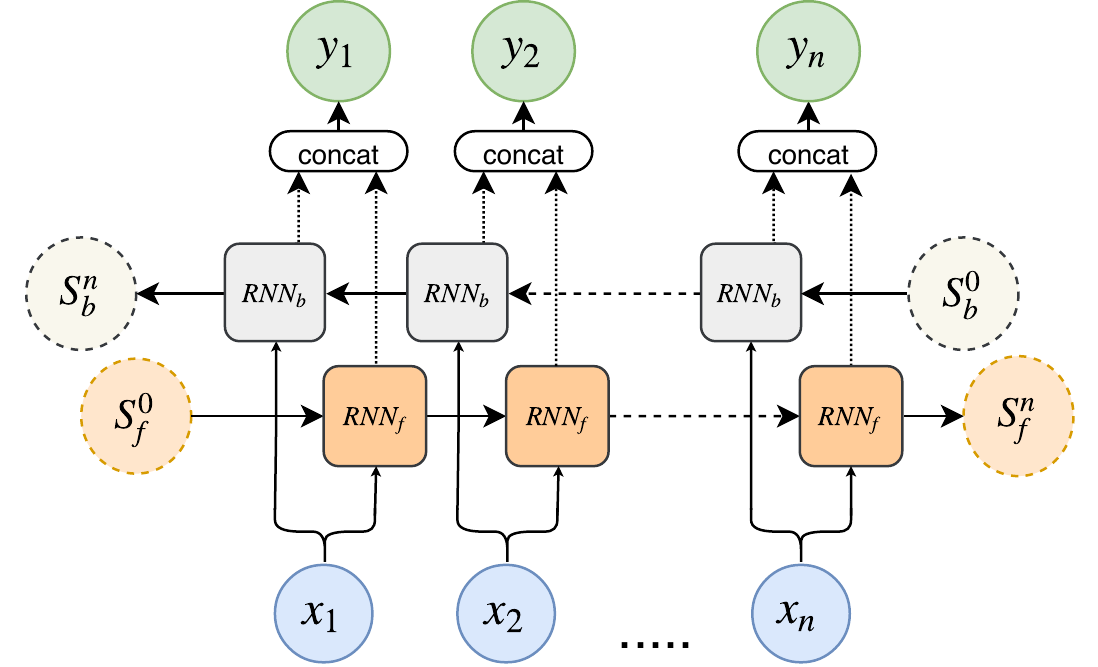}
	\vspace{-8pt}
	\caption{BiRNN scheme where at every index, the outputs of both forward and backward RNNs are concatenated.}
	\label{fig:bi_rnn}
\end{figure}

Bidirectional RNN (BiRNN) \cite{schuster1997bidirectional} is a model composed of two recurrent networks: the first one reads the input regularly while the later reads it backward. In our model, the outputs of both forward and backward RNNs are concatenated as depicted in \figref{fig:bi_rnn}. Such a model can predict every frame based on both past and future frames. 
A common approach to address a segmentation task is to locate the frames in which a new segment begins. To make this search tractable, we decompose $\sy$ into parts and score every part individually using a feature function. The global score is then the summation of these part scores, $\Phi(\sx, \sy) = \sum_{i=1}^{k}{\phi(\sx,y_i)}$. We use BiRNNs to learn the appropriate feature functions, capitalizing on their ability to focus in on the relevant features of the dynamic temporal sequence. 
We can now re-formulate our prediction rule in \eqref{eq:yw} as: \vspace{-3pt}
\begin{equation}
\begin{aligned}
\label{eq:dec_phi}
\syh_{\vw}(\sx) 
&= \argmax_{\sy \in \Yc^k} ~ \vw^\top \Phi(\sx, \sy) \\
&= \argmax_{\sy \in \Yc^k} ~ \vw^\top \sum_{i=1}^{k} \vphi(\sx, y_i) \\
& = \argmax_{\sy \in \Yc^k} ~ \vw^\top \sum_{i=1}^{k} \text{BiRNN}(\sx, y_i) \\
& = \argmax_{\sy \in \Yc^k} ~ \vw^\top \sum_{i=1}^{k} \text{RNN}_{f} (\sx, y_i) \overset{concat}{\oplus} \text{RNN}_{b} (\sx, y_i) 
\end{aligned}
\end{equation}
The notation $\text{RNN} (\sx, y_i) $, is the RNN's output at index $y_i$ for a given sequence $\sx$. We use gradient-based methods to minimize the objective \eqref{eq:reg-loss}, since both the RNN and the structural hinge loss are differentiable with respect to the input and the model parameters.
\subsection{VOT Classifier}\label{ssec:classifier}
\begin{figure}[t]
	\centering
	\includegraphics[height=4.2cm]{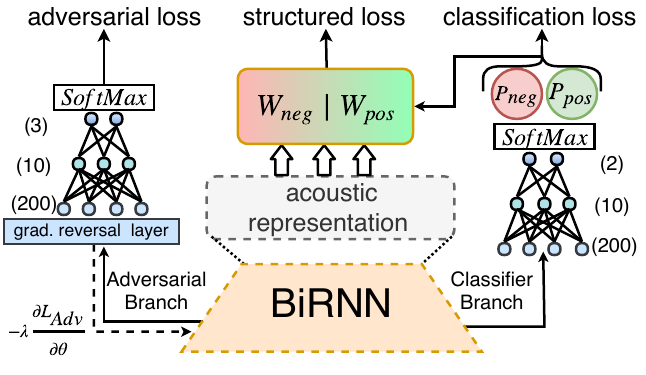}
	\vspace{-9pt}
	\caption{Our structured prediction model combining both classification and adversary branches.}
	\label{fig:model}
	\vspace*{-5.8pt}
\end{figure}

As noted above, VOT can be either negative or positive. Consider our prediction rule in \eqref{eq:yw}, the same weight vector $\vw$ is used regardless of the VOT type.
Having a VOT classification model, which tags whether the utterance has positive or negative VOT may improve our prediction. Specifically, use two different weight vectors: $\vw_{neg}$ when dealing with a negative VOT, otherwise, use $\vw_{pos}$. Rather than adapting two separate models (VOT predictor and VOT classifier), we integrate the tagging task with the main prediction. 
We do this by using an MTL technique where VOT tagging as positive or negative is the auxiliary task. Practically, the last hidden state of the RNN is forwarded to a small fully connected network, ending with a softmax layer of size two as depicted in \figref{fig:model}. The output of the softmax is a 2-valued vector which represents the probabilities for positive and negative, [$P(pos|\sx),P(neg|\sx)$]. The weight vector in \eqref{eq:dec_phi} is then set to be the one associated to the higher probability:
\begin{equation}
\vspace{-3pt}
\begin{aligned}
\vw := \left.
\begin{cases}
	\vw_{pos} &~~ P(pos|\sx) >  P(neg|\sx) \\
	\vw_{neg} &~~ $else$\\

\end{cases}
\right .
\end{aligned}
\end{equation}
The classifier parameters are optimized by minimizing the negative log-likelihood loss. As shown in \figref{fig:model}, we allow the gradients to backpropagate from the classifier and update the BiRNN parameters as well; thus, the tasks are jointly trained.

\subsection{Adversarial Multi-Task Branch}\label{ssec:sdversarial}

During training, a model will learn to map the features of seen examples to the desired outcome. However, if the seen examples do not include all the possible dynamics and behavior of the features, performance may be degraded on a different dataset with a different distribution over the features. 
To make our model robust in the face of this issue -- to make our model dataset-invariant -- we must develop a method by which the model can apprehend the relevant features while ignoring the rest. Conceptually, the RNN should ``filter'' the acoustic representation, removing those features that are not relevant to the task. 

To implement this, we feed the last state of the RNN to another small fully connected network which acts as a dataset classifier. This is similar to the VOT tagger in the previous section but has a different impact on the RNN. To encourage the main model to extract only those features which are task-related -- to avoid extraction of features that alter across datasets -- we add a gradient reversal layer \cite{ganin2014unsupervised} before feeding the adversarial net. This layer allows the normal flow of information in the forward step, but in the backward step, it changes the sign of the gradients. Hence, while the adversarial net tries to minimize its loss in dataset classification, the feature extractor is tuned to maximize this loss. At inference, the adversarial branch is irrelevant and can be ignored. 

Overall our model's loss, $\bar{\ell}_\mathrm{model}$ is given by:\vspace{-1pt}
\begin{equation}
\label{model_loss}
\bar{\ell}_\mathrm{model} =  \bar{\ell}_\mathrm{struct} +\bar{\ell}_\mathrm{tagger} +\lambda \bar{\ell}_\mathrm{adversarial},
 \end{equation}
where $\bar{\ell}_\mathrm{struct}$ is the structured max-margin loss defined in Eq.~\ref{eq:structured-loss}, $\bar{\ell}_\mathrm{tagger}$ is the NLL loss of the tagger,  $\bar{\ell}_\mathrm{adversarial}$ is the NLL loss of the adversarial branch, and $\lambda$ is a hyper-parameter to control the weight of the adversarial loss in the training objective.


\section{Experiments}
First, we describe the datasets and then proceed by comparing the performance of each model on both seen and unseen datasets to assess generalization ability.

\subsection{Datasets} In order to estimate a model's cross-datasets performance, we used four corpora of isolated productions of words with initial voiced and voiceless stops, characterized by different speakers and recording environments.
The first dataset named PGWORDS \cite{Paterson2011} consists of English picture naming and reading aloud by L1 English speakers and L1 Portuguese/L2 English bilinguals from the Northwestern University community. The other three (\cite{goldrick2013effects}, \cite{dmitrievaDataset}, \cite{Shultz2011,Shultz2012}) consist of recordings of single words read aloud by L1 English participants. The corpus described in \cite{goldrick2013effects} is composed of recordings from the Northwestern University community where \cite{dmitrievaDataset} and \cite{Shultz2011, Shultz2012} are recordings of speakers from the Purdue University community. Each corpus was divided into train, validation, and test; each speaker was uniquely assigned to one of these sets, allowing us to ensure that performance generalizes across talker differences.
The input features are described in detail in \cite{sonderegger2012automatic} and have been utilized in previous research \cite{adi2015vowel,adi2017sequence}. Broadly speaking, these features measure small time scale (1 and 5 msec windows) spectral, duration, and amplitude measures utilized by human annotators. Overall we have 63 features per frame. Among the entire training corpus consisting of 7,483 utterances, the ratio of positive to negative VOT is 80:20 respectively. 

\subsection{Evaluation and discussion}
We started by evaluating our proposed model \emph{Dr.VOT} compared to previous known models, namely \emph{AutoVOT} \cite{sonderegger2012automatic}, \emph{DeepVOT} \cite{adi2016automatic}, and \emph{DeepSegmentor} \cite{adi2017sequence}). In all the experiments we used a 2 layers BiLSTM \cite{hochreiter1997long} with a hidden size of 100 for each network, as the feature function,  and optimized parameters with $Adagrad$ \cite{duchi2011adaptive} and a learning rate of $0.01$; $\lambda$ was set to $0.1$. \tableref{tab:natalia_results} summarizes the performance achieved by each model as reported in previous papers \cite{adi2016automatic,adi2017sequence}, when trained and tested on the PGWORDS dataset. The performances were measured by the proportion of differences between automatic and manual measures falling at or below a given tolerance value. As noted in the introduction, the resolution needed for this task is quite fine-grained; we therefore focus primarily on the 2, 5, and 10 msec tolerances in order to reproduce phonological and clinical research. For example, for \emph{DeepSegmentor}, the difference between automatic and manual measurements in the test set was 2 msec or less in 78.2\% of examples. As one can see, our model, which is denoted as \emph{Dr.VOT}, performed almost identically to the state-of-the-art model \emph{DeepSegmentor}. 

\begin{table}[h]
	\small
	\renewcommand{\arraystretch}{1.3}
	\centering
	\footnotesize
	\caption{Proportion of differences between automatic and manual measures falling at or below a given tolerance value.} 
	\vspace{-5pt}
	\label{tab:natalia_results}  
	\resizebox{\linewidth}{!}{%
		\begin{tabular}{l|c|c|c|c}
			\hline
			\hline
			Model & $\tau\le$2 msec & $\tau\le$5 msec & $\tau\le$10 msec & $\tau\le$15 msec  \\
			\hline
			$AutoVOT$ & 49.1 & 81.3 & 93.9 & 96.0 \\
			\hline
			$DeepVOT$ & 53.8 & 91.6 & 97.6 & 98.7 \\
			\hline
			$DeepSeg.$ & 78.2 & 94.1 & 97.0 & 98.6 \\
			\hline
			$Dr.VOT$ & \bf{78.5} & \bf{94.6} & \bf{98.0} & \bf {98.8} \\
			\hline
			\hline
	\end{tabular} }
\end{table}

Using the previous models ``in the wild,'' i.e., on an entirely new dataset with different talkers from which the model was trained on, and a new acoustic environment, they face a severe performance degradation. For example, when testing \emph{AutoVOT} and \emph{DeepSegmentor} on the unseen corpus \cite{goldrick2013effects}, the results in the $\tau \leq 2$ msec section; the percentage of files in which the predicted VOT was off by up to 2 msec, dived from 49.1 to 31.9 and from 78.2 to 46.8, respectively. 

To better understand the effect of each component of our proposed model, we compared the following four models:
\renewcommand{\labelenumi}{\roman{enumi}}
\begin{enumerate}
    \item \emph{DeepSegmentor} \cite{adi2017sequence};
    \item Tagging model, denoted with \emph{TGM}, an MTL extension of \emph{DeepSegmentor} with the VOT classifier;
    \item Adversarial model, denoted with \emph{ADM}, is an adversarial side-branch added to \emph{DeepSegmentor};
    \item Both tagging and adversarial models, denoted as \emph{Dr.VOT}.
\end{enumerate}

	
	
To check the performance on unseen datasets, we used a 4-fold cross validation method on the datasets. In each fold one of the 4 datasets was left out for evaluation while the rest were used for training. Each model was evaluated on unseen examples from the same dataset it was trained on as well as on examples from the evaluation dataset. We balanced the data proportionately giving a higher probability to examples of negative VOT. Those probabilities were calculated at the beginning of each run to ensure the training set was equally distributed. We report in \tableref{tab:Loss} the average performance achieved by each model using the task loss in \eqref{eq:task_loss}.

\begin{table}[t]
	\small
	\renewcommand{\arraystretch}{1.3}
	\centering
	\footnotesize
	\caption {Proportion of differences between automatic and manual measures falling at or below a given tolerance value. Performance at each tolerance level is separated by within-corpus (first column) vs. unseen corpora (second column).} 
	\vspace{-5pt}
	\label{tab:Loss} 
	\resizebox{\linewidth}{!}{%
		\begin{tabular}{l|c|c|c|c}
			\hline
			\hline
			Model & $\tau\le$2 msec & $\tau\le$5 msec & $\tau\le$10 msec & $\tau\le$15 msec \\
			\hline
			$DeepSeg.$ & 64.8 $~\vert~$ 50.8 & 86.2 $~\vert~$ 78.8 & 93.6 $~\vert~$ 89.5  & 96.3 $~\vert~$ 93.2 \\
			\hline
			$TGM$      & 70.9 $~\vert~$ 53.8 & \textbf{90.7} $~\vert~$ 81.3 & \textbf{96.5}  $~\vert~$ \textbf{91.5} & \textbf{98.0} $~\vert~$ 94.6\\
			\hline
			$ADM$      & 69.1 $~\vert~$ 55.4 & 89.4 $~\vert~$ 81.1 &   95.8 $~\vert~$ 90.9 & 97.8 $~\vert~$ 94.2 \\
			\hline
			$Dr.VOT$ & \textbf{71.1} $~\vert~$ \textbf{56.5} &  90.3 $~\vert~$ \textbf{82.5} &   96.2 $~\vert~$ \textbf{91.5} & \textbf{98.0} $~\vert~$ \textbf{94.7} \\
			\hline
			\hline
	\end{tabular}   } 
\end{table}

The results suggest that the proposed model, $Dr.VOT$, handled untrained items from both seen and unseen corpora with greater accuracy than \emph{DeepSegmentor} at all tolerance values. At the lowest tolerance values on the trained dataset, both of our additions improve performance relative to the benchmark (64.8\%). Adding the tagging task via MTL had a slightly better impact (70.9\%) than adversarial training (with dataset classification as the adversarial task: 69.1\%). Combining both methods barely disrupts the model with respect to the tagger alone.
When looking at the reported results on an unseen corpus (the second of two columns at each tolerance value), the benchmark consistently underperforms the other models. Adding an adversarial branch contributes more to the performance at the lowest range of tolerances (from 50.8\% to 55.4\%) compared to the classification branch (from 50.8\% to 53.8\%). Across levels of tolerance, the \emph{Dr.VOT} exhibits the highest level of generalization.

When splitting a dataset into training, validation, and test sets, each set should represent the the distribution of the entire dataset. Consequently, results on a test set should be similar to performance in the ``real-world.'' As mentioned before, a primary goal of this paper is to address a key challenge when measuring VOT: producing accurate measurements when the VOT is negative. As we break down the reported results in \tableref{tab:Loss} to the performance on negative VOT utterances, we can clearly see the contribution of our proposed algorithm. Since the positive case is far more common, the results on the positive VOT examples are similar to those reported in \tableref{tab:Loss}. Hence, our discussion focuses on predictions for negative VOT. 

\begin{table}[t]
	\small
	\renewcommand{\arraystretch}{1.3}
	\centering
	\footnotesize
	\caption {For negative VOT examples from unseen corpora only, proportion of differences between automatic and manual measures falling at or below a given tolerance value.} 
	\vspace{-5pt}
	\label{tab:neg_loss} 
	\resizebox{\linewidth}{!}{%
		\begin{tabular}{l|c|c|c|c}
			\hline
			\hline
			Model & $\tau\le$2 msec & $\tau\le$5 msec & $\tau\le$10 msec & $\tau\le$ 15 msec \\
			\hline
			$AutoVOT$ & 3.7 & 7.9 & 14.2 & 20.2 \\
			\hline
			$DeepSeg.$ & 23.2  & 46.1  & 63.4 & 71.4 \\
			\hline
			$Dr.VOT$ & \textbf{32.4}  &  \textbf{55.3}  &   \textbf{75.7}   & \textbf{84.5} \\
			\hline
			\hline
	\end{tabular}   } 
\end{table}
The results for examples from unseen corpora are summarized in \tableref{tab:neg_loss}. It is clear that it was more difficult to predict negative VOT than positive VOT. This may be due to difficulties in consistent annotation of the onset of voicing, given that this often has extremely small amplitude in negative VOT. However, even considering the overall lower performance, it is clear that our model exhibits dramatic improvement over existing approaches. In 32.4\% of the utterances which have negative VOT, our model prediction was off by at most 2 msec, a 9.2\% absolute improvement and a roughly +40\% relative improvement to the 23.2\% achieved by the current state-of-the-art. When allowing larger threshold values, our method consistently showing a relative improvement of around +20\%. (Note the \emph{AutoVOT} model is limited to the positive case only, leading to very poor performance on the negative-VOT utterances.) This dramatic performance improvement on negative VOTs accounts for most of the overall performance boost of the current approach. For example, considering all VOTs, at a 10 msec threshold \emph{Dr.VOT} outperforms \emph{DeepSegmentor} by 2\% (91.5\% over 89.5\%); focusing only on the less common negative VOTs, there is a 12.3\% gain in performance (75.7\% over 63.4\%).

Finally, in addition to producing better predictions of segmentations, our model also successfully classifies whether the VOT is negative or positive with an accuracy of 94-97\% (depending on the dataset) on both seen and unseen corpora.

	\section{Conclusions}
We have presented two different approaches for generalization in a deep learning structured prediction framework. Focusing on enhancing overall performance, we combined MTL within the prediction rule. Focusing on generalization, we used adversarial learning to ``filter'' unnecessary properties from the feature space. As far as we are aware of, this is the first work to tackle, in an integrated way, VOT classification along with prediction of both positive and negative VOTs -- critically, yielding 
well-balanced results on both. In future work we would like to further explore the adversarial effect by testing it over multilingual corpora, and add handcrafted auxiliary related tasks. Such an approach may allow the first-ever application of these techniques to the analysis of bilingual language production.

After acceptance $Dr.VOT$ will be publicly available at {\small\url{https://github.com/MLSpeech/Dr.VOT}}.

\vspace{-5pt}

	\section{Acknowledgements}
	\vspace{-1pt}
	Research supported in part by NIH grant 1R21HD077140. Thanks to Olga Dmitrieva and Alexander Francis for providing access to VOT datasets.
	
	\bibliographystyle{IEEEtran}
	
	\bibliography{refs}
	
\end{document}